\documentclass{iaus}
\usepackage{graphicx}

\newcommand{\reff}{$r_{\mbox{eff}}$}
\newcommand{\mage}{$\langle\mbox{age}\rangle$}

\title[Radial Gradients in SDSS Disk Galaxies] 
{Radial Dependency of Stellar Population Properties in Disk Galaxies from SDSS Photometry}

\author[C.~W.~Yip and R.~F.~G.~Wyse]   
{Ching-Wa~Yip\thanks{Keck Fellow; cwyip@pha.jhu.edu;
    wyse@pha.jhu.edu.} and Rosemary F.~G.~Wyse}

\affiliation{Department of Physics and Astronomy, The Johns
  Hopkins University, Baltimore, MD 21218, USA  \\[\affilskip]}

\pubyear{2006}
\volume{241} 
\pagerange{xxx}
\date{?? and in revised form ??}
\jname{Stellar Populations as Building Blocks of Galaxies}
\editors{R.~F.~Peletier, A.~Vazdekis, eds.}
\begin{document}

\maketitle

\begin{abstract}
The   evolution  of  stellar   disks  is   of  great   importance  for
understanding  many aspects  of  galaxy formation.   In  this work  we
perform stellar  population synthesis on  radially resolved photometry
of 564 disk galaxies from the  SDSS DR5, selected to have both spectra
of  the  central  regions   and  photometry.   To  explore  fully  the
multi-dimensional likelihood space defined by the output parameters of
the spectral  synthesis, we use  Markov Chain Monte Carlo  to quantify
the expectation values, the  uncertainties and the degeneracies of the
parameters.  We  find  good  agreement between  the  parameter  values
obtained   using  the   SDSS   broad-band  colors   and  the   spectra
respectively. In general the derived mean stellar age and the best-fit
stellar metallicity  decline in  value from the  galaxy center  to the
outer  regions (around  1.5  half-light radii),  based on  sub-samples
defined  by  concentration  index.   We  also  find  that  the  radial
dependency of the stellar population parameters exhibits a significant
variation, and this diversity is  likely related to morphology and the
physics  of star  formation.  \keywords{galaxies:  stellar  content --
techniques: photometric}
\end{abstract}

\firstsection 
\section{Introduction}

The   quantification  of  radial   gradients  of   stellar  population
parameters provide  a powerful constraint on  galaxy formation models,
as distributions of stellar  metallicity and age are basic predictions
(e.g.~\cite{Larson76},    \cite{Wyse89},   \cite{Robertson04}).    The
completion of  the SDSS$-$I \cite{SDSS} has allowed  us to investigate
this issue with an unprecedented large sample of nearby disk galaxies,
with uniform photometric calibrations.

\section{Analysis and Main Results}\label{section:mainresults}

Our  sample  is chosen  from  the SDSS  DR5  disk  galaxies with  both
spectroscopic (for the central 3 arcsec) and photometric observations.
The availability  of both types of  data allow us  to characterize the
parameter   uncertainties  using  only   the  photometry.    They  are
low-inclination (isophotal  minor axis/major axis  = $0.25 -  1$) with
distance  ranging from  $\sim 20  - 700$~Mpc.   We adopt  the P\'EGASE
\cite{pegase} stellar  population synthesis  models, with the  ages of
the  oldest stars in  the range  from $12.5  - 13.7$~Gyr,  the stellar
metallicity from  $0.0001 -  0.05$, an exponential  star-formation law
with the e-folding time from  $0.2 - 12$~Gyr, and the reddening E(B-V)
from  $0.0 -  0.8$.  We  are thus  solving simultaneously  for several
parameters.   To obtain  estimates  and covariance  matrices of  these
parameters from  the SDSS radially  resolved photometry ($u,g,r,i,z$),
we developed  a Markov  Chain Monte Carlo  (MCMC) code to  perform the
minimization in  N-dimension. Each annulus  of each galaxy  is treated
independently in the minimization.

\begin{figure}[ht]
\begin{minipage}[t]{7cm}
\begin{center}
\includegraphics[width=6.8cm,clip]{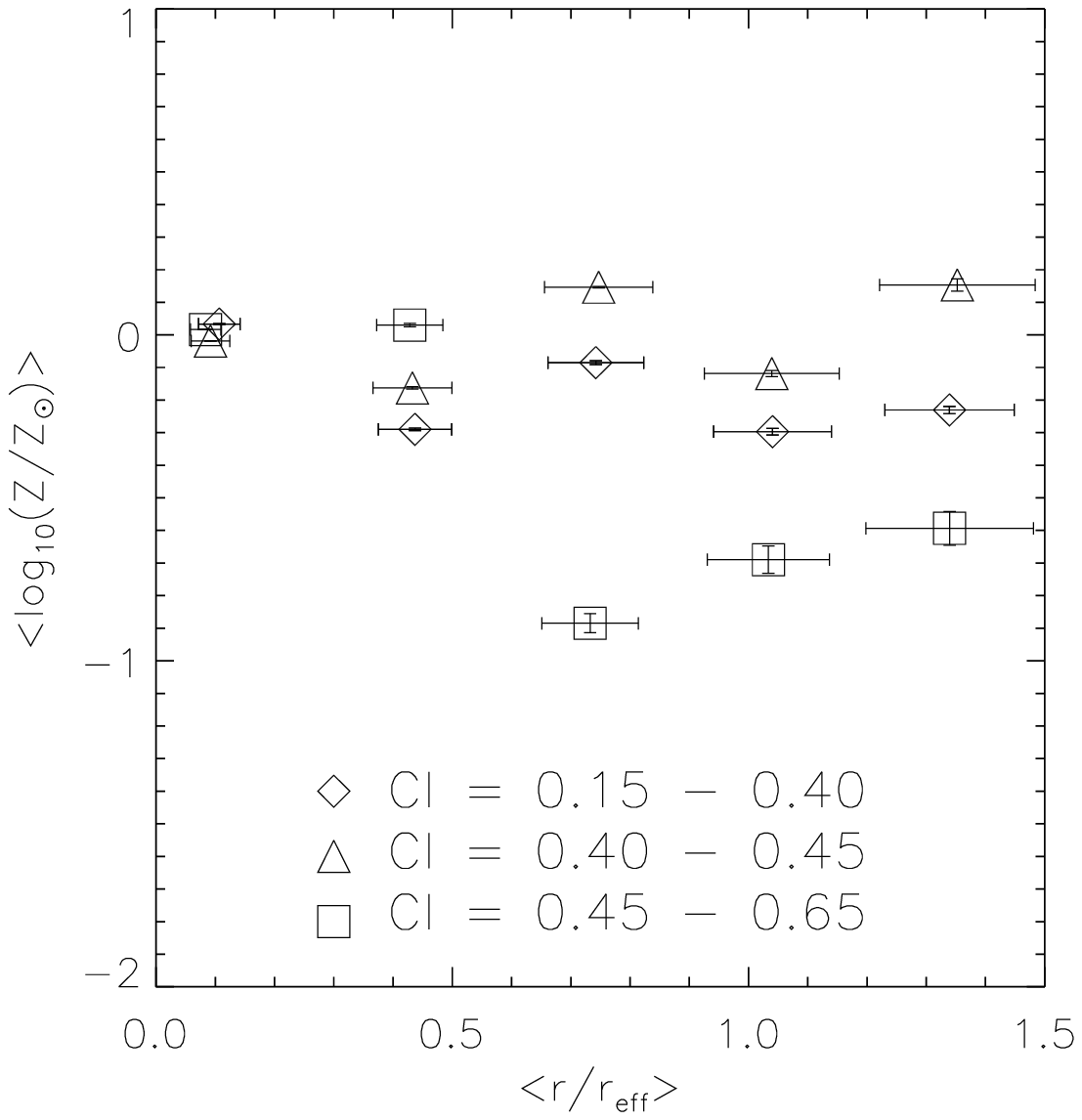}
\caption{Radial  dependency  of stellar  metallicity,  averaged  in
  three  ranges of  concentration index  (CI):  $0.15-0.40$ (rhombus),
  $0.40-0.45$ (triangle) and $0.45-0.65$ (square). The smaller the CI,
  the more concentrated is the galaxy light.}
\label{fig:yip_2_fig1}
\end{center}
\end{minipage}
\hfill
\begin{minipage}[t]{7cm}
\begin{center}
\includegraphics[width=6.8cm,clip]{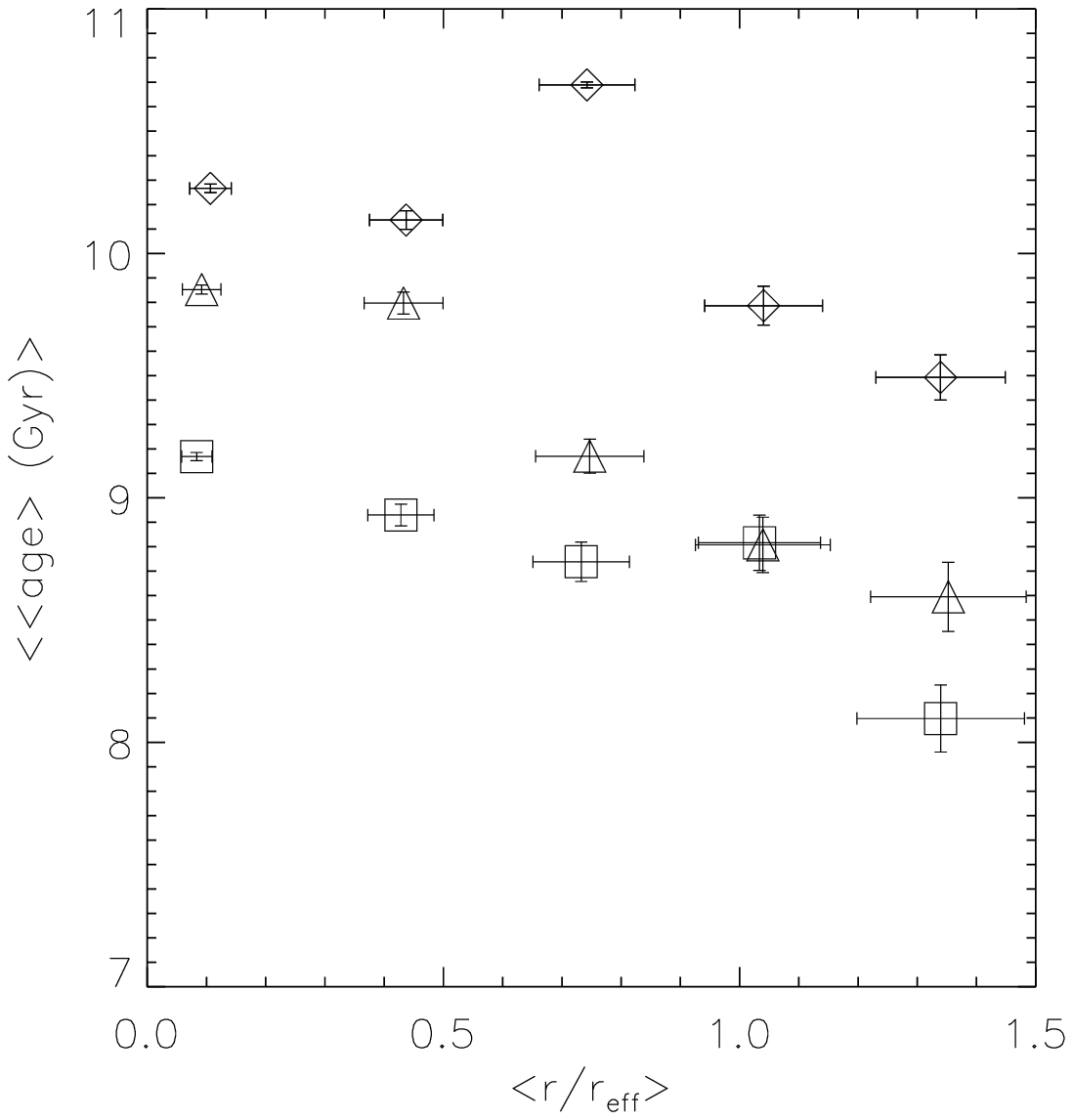}
\caption{Radial dependency of the mean stellar age.}
\label{fig:yip_2_fig2} 
\end{center}
\end{minipage}
\end{figure}

The  stellar  metallicity  and  \mage   \  as  a  function  of  radius
($r/$\reff, where \reff \ is  the half-light radius of the galaxy) are
shown    in    Fig.~\ref{fig:yip_2_fig1}   and    \ref{fig:yip_2_fig2}
respectively.  The error bar indicates  one sigma of the mean value in
each axis. Radial dependencies are  found for these two parameters, in
that they  decrease toward larger  radii (cf.~similar results  for 121
disk galaxies,  from a combination  of broad-band optical and  near IR
data  from \cite{Bell00}).  In  addition, we  find a  clear separation
between galaxies of  different $r-$band concentration indices (CI's)):
the more concentrated  are the galaxies, the higher  the values in the
stellar  metallicity   and  \mage.   While  the   overall  trends  are
compatible with  `inside-out' models  of disk galaxy  formation, there
are  interesting systematics in  the gradients  shown in  the figures.
Further, we see features  in individual galaxies at intermediate radii
that could be  a manifestation of resonances associated  with bars and
we are embarking on  a more detailed morphological characterization of
the  galaxies  e.g.~bulge  to  disk  ratio,  barred  nature  etc.   In
particular, some  models predict that  bars could lead to  a localized
old   age   within   $r/$\reff~$\sim  0.7$   (V.~Debattista,   private
communication).   The details  of  our analysis  and  results will  be
presented in a separate paper.

We acknowledge support from the W.~M.~Keck Foundation, through a grant
given to establish a program of data intensive science at JHU.

\end{document}